# ON THE USAGE OF DATABASES OF EDUCATIONAL MATERIALS IN MACEDONIAN EDUCATION

Mimoza Anastoska-Jankulovska and Pece Mitrevski

Faculty of Information and Communication Technologies,
University "St. Kliment Ohridski", Bitola, Republic of Macedonia

## ABSTRACT

*Technologies have become important part of our lives. The steps for introducing ICTs in education vary from country to country. The Republic of Macedonia has invested with a lot in installment of hardware and software in education and in teacher training. This research was aiming to determine the situation of usage of databases of digital educational materials and to define recommendation for future improvements. Teachers from urban schools were interviewed with a questionnaire. The findings are several: only part of the interviewed teachers had experience with databases of educational materials; all teachers still need capacity building activities focusing exactly on the use and benefits from databases of educational materials; preferably capacity building materials to be in Macedonian language; technical support and upgrading of software and materials should be performed on a regular basis. Most of the findings can be applied at both national and international level – with all this implemented, application of ICT in education will have much bigger positive impact.*

## 1. INTRODUCTION

A *Database of Educational Material* (DEM) is an organized collection of digital educational materials: sometimes it has free access, sometimes it can be updated by the user. Regardless of that, its main aim is to provide materials that can facilitate the educational process by using the existing technologies. Structured DEMs are the next developmental stage; they consist of more information about the implemented changes, their author, usage of material, achieved results, time and date of accomplishments, etc. Using structured DEMs in education will provide more information about the progress of students, but will also provide teachers with an opportunity to design own subset of materials that is suitable for them and their students.

During the previous decade in schools in the Republic of Macedonia a lot of hardware has been installed, and teachers have received a lot of capacity building training: computers in schools are installed and connected on the Internet; numerous software applications have been translated and adopted for Macedonian education; each teacher has received a laptop to facilitate own preparation for work and implementation of teaching activities [1]. Having in mind that for more than a decade different projects were focusing on preparation of digital teaching/learning materials for Macedonian schools, it is important to see *if* and *how* the digital materials have changed the educational process, and *if* and *how* the approach of teachers has been changed.

The problem targeted with this research was exploring *how* databases of digital educational materials are used and *how* they are affecting the education in urban schools in the Republic of Macedonia. Research's aim was to identify to what extent structured databases of educational materials are used in Macedonian educational system. After a critical engagement with relevant literature in Section2, we explain the research protocol and methods in Section 3 and present the research findings in Section 4, whereas Section 5 concludes the paper.

## 2. RELATED WORK

Generally, the usage of ICT in education is a relatively new topic all across the world. Possibilities of introducing the new technologies in classes are widening and changing every day. Researches have been done in some countries with more advanced usage of ICTs. One important finding that reappears is that transition from traditional teaching and learning into electronic will not change anything if same strategies, methods and techniques are continued to be implemented [2]. This is also supported by an OECD finding [3] that crucial improvements were not visible in the countries that have made huge investments in ICT in schools. The first recommendation internationally is that together with introducing new technologies in schools, changes in approaches and practices should be introduced.

Another finding is that vocational teachers are not prepared enough for implementing ICT in pedagogy, which results in divided approaches in schools [4]. Another recommendation emerging from several researches is that teachers should be adequately prepared and their capacity should be continuously built in order to establish and maintain the quality results from implementation of ICT in schools. Teachers should be agents of change in schools, not only with their acceptance and introduction of new technologies, but with creating and designing of the technological innovations [5].Teachers should be leading not only the implementation, but also the creation of digital educational materials in schools.

## 3. RESEARCH PROTOCOL AND METHODS

The guiding hypothesis of this research was "Implementation of DEMs promotes bigger quality in education through more motivated teachers and better learning results". This research has been implemented in urban schools across the country, not selecting any educational level in particular. In total 182 teachers were interviewed, and their answers were analyzed. The questionnaire was done as short as possible, consisting of 17 questions in total. Few general questions in the beginning of the questionnaire were about gender, type of school, or working experience in education. These questions were included in order to define if there is correlation of these data to usage of databases of educational materials. Then questions about the use of ICT in education, for work and private purposes followed. There were several open-ended questions at the end of the questionnaires aiming to collect more detailed information about what kind of materials and software are used by teachers, and also to receive more qualitative data about the teachers' approach towards ICT. These questions were not compulsory – if a teacher did not have any experience in the topics, he/she could leave it unanswered. These questions were increasing the test's validity – even an unanswered question is giving information for the analysis. With this approach, additional information about teachers' experiences was received.

Statistical analysis of received answers has been performed. Received results are showing that valid instrument was used. Cross-tabulating, as a tool for defining the frequency of answers with the combination of two or more questions, was one of the analytical approaches used. By combining the answers from two questions, more insight about the teachers' views and approaches towards databases of educational materials is received.

The distribution of the respondents across gender, age, or schools' levels is matching the general distribution of teachers against these criteria in all schools in the Republic of Macedonia; hence, it is supporting the reliability of the research and endorsing the findings from this research. The selected finding from this research is presented in this paper and is illustrating the most characteristic research findings.

## 4. RESEARCH FINDINGS

Considering the availability of hardware in schools and laptop computers for teachers, the questionnaire for this research was uploaded online with the aim to ease access to it, and give teachers the possibility to fill it at preferred place and time. At the commencement point of the research, even before starting the data processing and the data analysis, one finding emerged: The teachers were not willing to fill online questionnaires. Although, the link to it and information about the research were distributed in many schools across the country, only 1/3 of all final respondents (60 teachers) have used this online possibility. However, when the teachers received paper copies of the questionnaire, their response rate was much higher. The reasons can be numerous:preference of human interaction while delivering paper copies; no willingness to work with digital documents; problems with the equipment in schools; or something else. This research had no ambition to define the reason behind this striking finding. The reason why teachers are not willing to fill online surveys/documents can be a topic for a future research.

The total of 182 teachers have responded on paper or online questionnaires. The gender distribution of respondents is: 32% male and 68% female. Their age distribution is as follows.

Fig. 1 presents the age distribution of respondents: 27% less or equal to 35 years old, 34% between ages of 36 and 45, 33% between ages of 46 and 55, and 6% more than 55 years. The gathered data from these questions were cross-tabulating with other answers to determine if there is a difference in understanding and approaches between genders or age groups, or experience with ICTs or the duration of experience in schools has bigger impact.

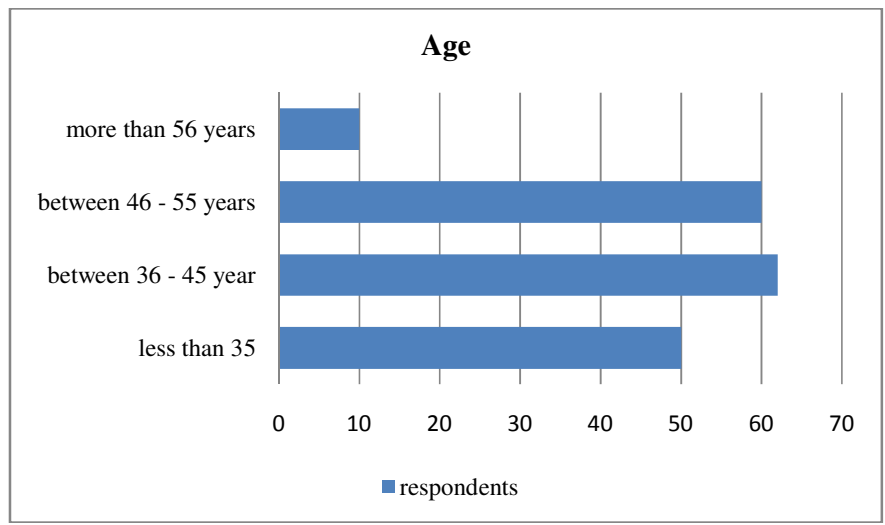

Figure 1. Age distribution of respondents

The predominant number of respondents have said that they are using ICTs for private purposes for more than 10 years (>64%), and the others between 3 and 10 years. Most of the respondents (64%) have stated that they are using ICTs in schools between 3 and 10 years. Usage of ICT in education is presented in Fig. 2.

Several questions were focusing on the databases of educational material (DEM). During the implementation of different projects and training activities in the past period, a lot of digital educational materials were developed and shared between different groups of teachers. It was expected that the interviewed teachers will have experience with DEM and can explain own understanding and approach.

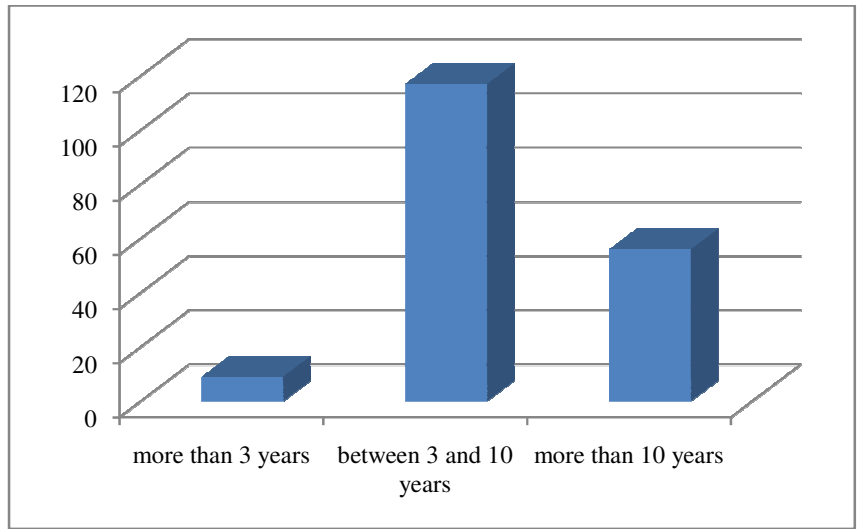

Figure 2. Distribution of respondents based on
duration of experience with use of ICT in education

The interesting result is that 15% of the teachers think that DEM will not affect or will reduce the success of students learning. Although this percent is not high, it is still significant since it is supporting the finding of international researches saying that only with introducing ICT and DEM in classes, instant success and change cannot be expected [6], [7]. With this first recommendation will be that *introduction of ICT in education should be done in the same time with introduction of new educational approaches by using these technologies.*

Important is the percent of teachers that don't have opinion, and with that meaning no experience with DEM, which is 24% (Fig. 3). This is showing that although a lot of technology is installed in schools, not all teachers have experience with using digital educational materials. Some future research should focus on why the technologies are not used to full extend in schools and why not all teachers have experience in implementing digital educational materials.

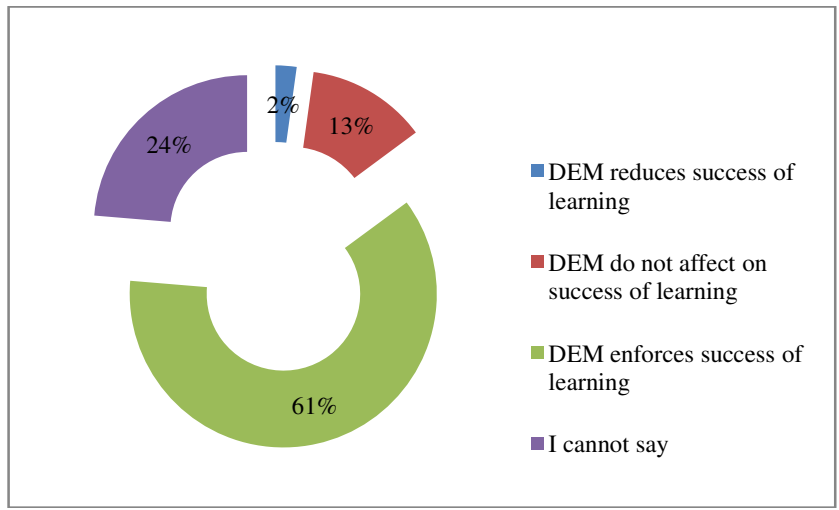

Figure 3. Effect of usage of DEM on success of learning

Big percent of the respondents (≈60%) have selected that they think DEM is helping them improve the effectiveness of their own work (Fig. 4). Approximately 30% have replied that they cannot say how DEM will affect their work, which means that they do not have enough experience in implementing DEM in education.Having in mind the previously mentioned fact that there have been a lot of materials developed, this finding is revealing that these materials are not widely used in schools. The recommendation will be*to make a collection with all previously prepared digital educational materials to be available for teachers to use in everyday work.*Guidelines can be drafted for new teachers tobe able to use all existing materials.

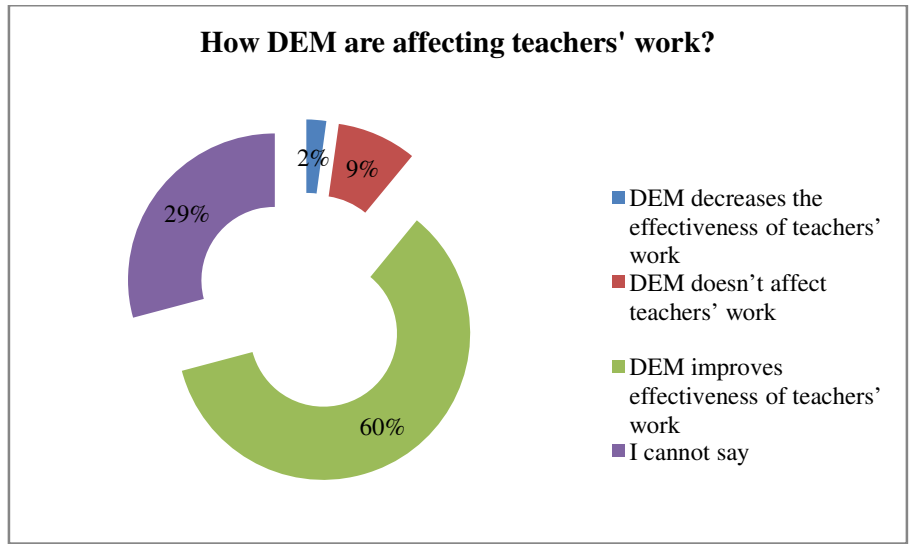

Figure 4. Effect of usage of DEM on effectiveness of teachers' work

Each question is shortened with abbreviation presented in the below tables. Questions are divided in two groups – independent variables/questions(Table 1) and dependent variables/questions (Table 2).

Table 1. Abbreviations for independent variables.

| Abbreviation | Question |
| --- | --- |
| **TU** | 2. What type is your school? |
| **POL** | 3. Your gender? |
| **GOBR** | 4. How long do you work in education? |
| **VOZR** | 5. How old are you? |
| **KOBR** | 6. How long are you using ICT in education? |
| **KPC** | 7. How long are you using ICT for private purposes? |
| **DNMOB** | 8. How long are you using digital educational materials in education? |

Table 2. Abbreviations for dependent variables.

| Abbreviation | Question |
| --- | --- |
| **POMR** | 10. According to you, to what extend are digital educational materials supporting teachers work? |
| **SOVL** | 11. According to you, to what extend databases of educational materials are supporting successful learning? |
| **PBDNM** | 12. How often would you like to use digital educational materials in your classes? |

Correlation (Comparative analysis) can be done with comparing answers to separate questions. To accomplish this, first answers are coded with numbers. Those questions with 4 possible answers can have numbers from 1 to 4, others with 3 options can have answers from 1 to 3 etc. Using coded answers like this, average of received answers, median, standard deviation, variance and other statistical data can be obtained. They are presented in thetable below (Table 3) for each of the questions.

Table 3. Statistical analysis for each of the questions.

| | Min | Max | Sum | No. of answers | Average | Median | Standard deviation | Sum of squares of stand. dev. | Variance |
|---|---|---|---|---|---|---|---|---|---|
| TU | 1 | 3 | 349 | 182 | 1.9176 | 2 | 0.8179 | 121.763 | 0.66903 |
| POL | 1 | 2 | 306 | 182 | 1.6813 | 2 | 0.8162 | 39.516 | 0.21712 |
| GOBR | 1 | 3 | 466 | 182 | 2.5604 | 3 | 0.8185 | 54.835 | 0.30129 |
| VOZR | 1 | 4 | 394 | 182 | 2.1648 | 2 | 0.8179 | 145.055 | 0.79700 |
| KOBR | 1 | 3 | 411 | 182 | 2.2582 | 2 | 0.8202 | 52.863 | 0.29045 |
| KPC | 1 | 3 | 482 | 182 | 2.6483 | 3 | 0.8184 | 43.494 | 0.23898 |
| DNMOB | 1 | 3 | 385 | 182 | 2.1154 | 2 | 0.8166 | 50.575 | 0.27789 |
| POMR | 1 | 4 | 575 | 182 | 3.1593 | 3 | 0.8147 | 80.379 | 0.44164 |
| SOVL | 1 | 4 | 558 | 182 | 3.0659 | 3 | 0.8142 | 81.209 | 0.44620 |
| PBDNM | 1 | 3 | 425 | 182 | 2.3352 | 2 | 0.8122 | 80.555 | 0.44261 |

These data can be used to determine correlation between received answers for each questions (Table 4). The smaller received number is (in absolute value), the smaller the connection between these two questions is. The bigger received number is (in absolute value), the bigger the connection between these two questions is. The biggest value is 1, meaning the direct relation between the two questions.

Table 4. Correlation between received answers to questions.

| | TU | POL | GOBR | VOZR | KOBR | KPC | DNMOB | POMR | SOVL | PBDNM |
|---|---|---|---|---|---|---|---|---|---|---|
| TU | 1 | | | | | | | | | |
| POL | -0.083 | 1 | | | | | | | | |
| GOBR | 0.189 | 0.075 | 1 | | | | | | | |
| VOZR | 0.244 | -0.032 | **0.731** | 1 | | | | | | |
| KOBR | 0.273 | -0.044 | **0.551** | 0.471 | 1 | | | | | |
| KPC | 0.353 | -0.082 | 0.182 | 0.171 | 0.407 | 1 | | | | |
| DNMOB | 0.213 | -0.073 | 0.441 | 0.333 | **0.591** | 0.285 | 1 | | | |
| POMR | **0.014** | -0.031 | 0.056 | 0.021 | 0.023 | **0.003** | -0.037 | 1 | | |
| SOVL | 0.020 | -0.056 | -0.041 | 0.037 | 0.105 | -0.097 | -0.084 | 0.236 | 1 | |
| PBDNM | 0.132 | -0.028 | 0.027 | -0.038 | 0.280 | 0.193 | **0.328** | -0.046 | **0.012** | 1 |

The important information can be derived from correlation between independent and dependent variables. In the above table that is presented with grey cells. In this part it can be noted that the biggest value is received in correlation between duration of usage of digital educational materials and desire to use DEM frequently and with all classes and subjects. The smallest is the correlation between type of school and motivation of teachers for usage of DEMs. Unexpectedly small is the correlation of usage of ICT for private purposes and motivation for usage of DEMs in education. This means that usage of ICT for private purposes is not affecting implementation of DEMs in classes, but some other factors are influencing it.

Several tables have been received with cross-tabulating of the answers. Selection of them is presented in the following text. Table 5 is presenting the frequency of answers on how the usage of DEM is affecting the teachers' work divided by gender of respondent.

Table 5. Cross-tabulation betweengender and effect of usage of DEM on teachers' work

| % | DEM decreases the effectiveness of teachers' work | DEM doesn't affect teachers' work | DEM improves effectiveness of teachers' work | I cannot say | TOTAL |
|---|---|---|---|---|---|
| Male | 0.0 | 10.3 | 60.3 | 29.3 | 100 |
| Female | 3.2 | 8.1 | 59.7 | 29.0 | 100 |

For both genders, the biggest percent of teachers think that DEM will improve the effectiveness of teachers work. However, it is important to note that for both gender this percent is almost the same. These data are not supporting the bias that one gender is more open to usage of technology; these data are showing that both gender are similarly open to usage of databases of educational materials in education in terms of improving effectiveness of teachers' work.

It is important to realize that there is no difference based on gender in acceptance of DEM and ICT in educational processes as a general rule. In all questions, the general distribution is almost equal between genders without any difference – in general, women are equally willing to use DEM in their work as men are.

A bigger percent of respondents (55.6%) with less than 3 year experience in use of computers cannot say how it will affect teachers' work, which is expected (Table 6). With increasing the experience of use of computers, the percent of the teachers that think that DEM will increase the effectiveness of teachers work. The future work on *capacity development of teachers should target use of DEM in education* in order to provide experience to teachers sooner in their working life, so that they can implement DEM more freely and benefit from their potential. The experience of teachers that were using DEM in education longer is very positive and is stating that also students can better learn the material by using DEMs [8].

Table 6. Cross-tabulation betweenduration of computer usage and perception of DEM's usage effect on teachers' work

| % | DEM decreases the effectiveness of teachers' work | DEM doesn't affect teachers' work | DEM improves effectiveness of teachers' work | I cannot say | TOTAL |
|---|---|---|---|---|---|
| <= 3 years | 0.0 | 22.2 | 22.2 | 55.6 | 100 |
| between 3 and 10 years | 3.4 | 12.0 | 53.8 | 30.8 | 100 |
| > 10 years | 0.0 | 0.0 | 78.6 | 21.4 | 100 |

If a teacher has short experience with DEM, he/she cannot decide how it will affect the success of learning. With the increase of length of teachers' experience with DEM, the percent of teachers thinking in favor of usage of DEM for successful learning is increasing. The biggest percent of teachers that think DEM is enforcing success of learning (almost 90%) is for teachers that have been using DEMs for more than 10 years (Table 7); this is also supported in literature and international research [9]. Helping and supporting teachers in introducing and implementing DEM as early as possible will result in quicker and more effective adoption and usage of DEM.

Table 7. Cross-tabulation between duration of DEM's usage and
perception of DEM's effect on students' learning

| % | DEM reduces success of learning | DEM do not affect on success of learning | DEM enforces success of learning | I cannot say | TOTAL |
|---|---|---|---|---|---|
| <= 3 years | 0.0 | 18.8 | 37.5 | 43.8 | 100 |
| between 3 and 10 years | 3.1 | 14.0 | 56.6 | 26.4 | 100 |
| > 10 years | 0.0 | 5.4 | 89.2 | 5.4 | 100 |

Teachers working in VET schools in the Republic of Macedonia are willing to implement DEM in their classes; and most of them want to do it with all classes and subjects. On the contrary, most of the teachers from gymnasia are willing to use DEMs only partially, with some classes or subjects (Table 8). It cannot be neglected the percentage of teachers from basic schools and from gymnasia that are not willing to use DEM at all. That may mean that they have no experience with DEM or their experience is negative. More research should be done in order to determine what the reasons for this percentage (13%) are.

Table 8. Cross-tabulation betweentype of school and
willingness to implement DEMs in education

| % | I do not want to use DEM | Only with some classes/subjects | I want to use DEM with all classes/subjects | TOTAL |
|---|---|---|---|---|
| basic education | 13.0 | 44.9 | 42.0 | 100 |
| gymnasium | 13.6 | 50.8 | 35.6 | 100 |
| VET | 5.6 | 37.0 | 57.4 | 100 |

Younger teachers are accepted as "digital natives", by default. Therefore, it was expected that younger teachers will be more eager in accepting ICTs and DEMs in their own work. However, findings are that older teachers and teachers with longer teaching experience are more willing to use DEM with all the classes and during all of the subjects. This can be observed also in the following Table 9, which is showing no difference in acceptance of DEMs between gender, but is striking that younger teacher are more reluctant in accepting DEMs in their everyday working practice. One of the reason is that younger teachers are not adequately prepared for using DEMs in their professional life [10].Future *capacity building activities should especially target younger teachers* too, without positive discrimination of them and prejudice that they are "digital natives", and that they have experience in how to use any technology under any circumstance.

The qualitative analysis of the open-ended questions has revealed several important findings. Experiences in implementing DEMs in educational process are divided – there are positive and negative ones. The given answers are supporting the previous quantitative finding that with only introducing DEMs without other additional changes, no positive result and improvement can be expected. Another finding is that the teachers still need a lot of training and capacity building in effective implementation of DEMs; permanent technical support should be available combined with timely upgrading of hardware and software. Another recommendation coming from open-ended questions is that larger volume of *digital materials in Macedonian language* will help and facilitate the introduction of DEMs at all levels.

Very positive was the finding that teachers understand the importance of the professional development and are doing their best to use all available resources, especially with the help of technology, to improve own development. This teachers' attitude towards professional development is important because it is also found that *teachers need quite a lot of training on understanding, using, preparing and updating DEMs*. This is especially important because a lot of the respondents have identified as weakness non-existing (or very few) digital educational materials in Macedonian language; in the near future, a lot of digital educational materials in Macedonian language should be developed. With this, *DEM in Macedonian language should be created*. This area should be in a focus of the future developments in education.

Table 9. Cross-tabulation between teachers' gender, teachers' age and
willingness to implement DEMs in education

| | | How would you like to use DEM? | | |
|---|---|---|---|---|
| | **How old are you?** | **I do not want to use DEM** | **Only with some classes/subjects** | **I want to use DEM with all classes/subjects** |
| male | <= 35 years | 2 | 7 | 6 |
| | between 36 and 45 years | 0 | 11 | 10 |
| | between 46 and 55 years | 3 | 5 | 9 |
| | >= 56 years | 2 | 0 | 3 |

| | | How would you like to use DEM? | | |
|---|---|---|---|---|
| | **How old are you?** | **I do not want to use DEM** | **Only with some classes/subjects** | **I want to use DEM with all classes/subjects** |
| female | <= 35 years | 0 | 21 | 14 |
| | between 36 and 45 years | 7 | 16 | 18 |
| | between 46 and 55 years | 5 | 19 | 19 |
| | >= 56 years | 1 | 2 | 2 |

# 5. CONCLUSIONS AND RECOMMENDATIONS

The experience of Macedonian teachers with DEMs is very various even within one single school. One of the most important recommendations from this research is that adequate capacity building should be implemented in order to reach similar level of knowledge and understanding for most teachers. All teachers should be trained to be able to design, implement and upgrade DEMs. Preparation of digital educational materials in Macedonian language is the prerequisite in order to improve the quality of education and to achieve the same level of technology usage in all schools and by all teachers and students.

Introducing only of DEMs in schools itself cannot result in big improvement because adequate attitude and practices should be in place, too. This research has identified that teachers should be equipped with adequate teaching approach in order to be able to use DEMs efficiently and to achieve the desired results. Another recommendation is to implement frequent periodic professional development training for all Macedonian teachers in order to exchange experiences and improve their own teaching approaches while using technologies and databases of educational materials.

This research was focusing on Macedonia as a specific case. However, its findings can be generalized on the international arena – continuous capacity building is important for every institution; teachers should be adequately prepared to develop, use and update DEMs; usage of materials in national language is prerequisite in every country; established practices and attitudes in schools should be targeted and revised in order to have efficient usage of DEMs.

Continuous capacity building of teachers will enable teachers to improve and adapt to the usage of DEMs. Updating DEMs and sharing experiences about it will support and help in producing new and update the existing DEMs. Materials in national language will be useful to all stakeholders in schools regardless of their age levels. But, most importantly, introducing DEMs itself will not change *anything* if practices are not adapted, too – this finding has already beenemphasized in international research and was identified with this research, as well.

# REFERENCES

[1]      Anastoska-Jankulovska, M. (2009) *Information systems in education*, Bitola, Macedonia.

[2]      World economic forum (2015)*Does technology in schools improve education?*,https://agenda.weforum.org/2015/09/does-technology-in-schools-improve-education/

[3]      OECD (2015) "Students, Computers and Learning: Making the Connection", PISA, OECD Publishing. http://dx.doi.org/10.1787/9789264239555-en

[4]      Uzunboylu, H., & Tuncay, N. (2010) "Divergence of Digital World of Teachers", *Educational Technology & Society*, 13(1):186-194.

[5]      OECD (2015) *Students, Computers and Learning: Making the Connection,* PISA, OECD Publishing. http://dx.doi.org/10.1787/9789264239555-en

[6]      Bingimlas, K.A. (2009) "Barriers to the Successful Integration of ICT in Teaching and Learning Environments: A Review of the Literature", *Eurasia Journal of Mathematics, Science and Technology Education*.

[7]      National Council for Curriculum and Assessment (2007) *ICT Framework – A structured approach to ICT in Curriculum and Assessment*, Ireland.

[8]      Scrimshaw, P., ed. (2004)*Enabling teachers to make successful use of ICT*, Coventry: British Educational Communications and Technology Agency, UK.

[9]      Tondeur, Jo, van Braak, Johan, Sang, Guoyuan, Voogt,Joke, Fisser, Petra& Ottenbreit-Leftwich, Anne (2012)"Preparing pre-service teachers to integrate technology in education: A synthesis of qualitative evidence", *Computers and Education*, 59(1):134-144.

[10]     West, D. (2013)*Mobile Learning: Transforming Education, Engaging Students, and Improving Outcomes*, Center for Technology Innovation in Brookings, Washington, USA.

## Authors

**Mimoza Anastoska-Jankulovska** is a PhD student at the Faculty of Information and Communication Technologies, University "St. Kliment Ohridski" – Bitola, Republic of Macedonia. She has extensive experience in teaching with and about ICTs. She has been pioneering introduction of ICT in educational process and has vast experience in teacher training for effective usage of ICT in educational process. Her research interest is e-learning and digital educational materials. 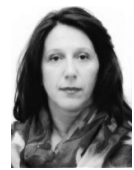

**Pece Mitrevski** received his BSc and MSc degrees in Electrical Engineering and Computer Science, and the PhD degree in Computer Science from the Ss. Cyril and Methodius University in Skopje, Republic of Macedonia. He is currently a full professor and Dean of the Faculty of Information and Communication Technologies, University "St. Kliment Ohridski" – Bitola, Republic of Macedonia. His research interests include Computer Architecture, Computer Networks, Performance and Reliability Analysis of Computer 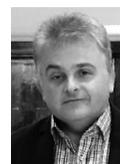 Systems, e-Commerce, e-Government and e-Learning. He has published more than 100 papers in journals and refereed conference proceedings and lectured extensively on these topics. He is a member of the IEEE Computer Society and the ACM.